\documentclass[twocolumn,showpacs,preprintnumbers,amsmath,amssymb,prb]{revtex4}
\usepackage{graphicx}% Include figure files
\usepackage{dcolumn}% Align table columns on decimal point
\usepackage{bm}% bold math

\begin{document}
%\draft

\title{Thermal van der Waals Interaction between Graphene Layers}

\author{ G. G\'omez-Santos}

\affiliation{Departamento de F\'{i}sica de la Materia
Condensada and Instituto Nicol\'as Cabrera, Universidad
Aut\'onoma de Madrid, 28049-Madrid,~Spain}

%\maketitle 

\begin{abstract} 
The van de Waals interaction between two graphene sheets is
studied at finite temperatures. Graphene's thermal length $(\xi_T = \hbar v /
k_B T)$ controls the force versus distance $(z)$ as a crossover from the zero
temperature results for $z\ll \xi_T$, to a linear-in-temperature, universal
regime for $z\gg \xi_T$. The large separation regime is shown to be a
consequence of the classical behavior of graphene's plasmons at finite
temperature. Retardation effects are largely irrelevant, both in the zero and
finite temperature regimes. Thermal effects should be  noticeable in the van de
Waals interaction  already for distances of tens of nanometers at room
temperature.
\end{abstract}


\maketitle 

\section {Introduction} \label{sec:intro}

Graphene, the single layer honeycomb lattice of carbon atoms
that forms graphite, has been realized experimentally in recent
times \cite{Novos04,Novos05}. Its  electronic properties characterized by a
linear dispersion around Fermi points, fixed by charge neutrality ({\it
massless}
Dirac fermions with velocity $v\sim 10^6 \; m/s$), have long attracted theoretical
interest \cite{Castro09}. But it is the present experimental accessibility, including Fermi level
tuning by gate voltages, what has unleashed an explosion of activity, fueled in
part by the prospects of tailoring  its electronic (and perhaps magnetic) 
properties in the nanoscale \cite{Geim07}.

More traditional areas like the van der Waals (vdW) interaction have also
benefited from the present interest. Although graphite is often characterized as
a vdW stack of graphene layers,  fundamental aspects such as the asymptotic
behavior of the vdW interaction between two graphene layers, have been unveiled
only recently by Dobson {\it et al.} \cite{Dobson06}, and shown not to conform
to the naive {\it sum of $R^{-6}$ contributions} \cite{Dobson06}. Taking as
reference the progress in accurate measurements of vdW interactions in general
\cite{Lamoreaux97}, the expected increase  in availability of graphene
\cite{Geim07,Geim09}, and its unique conceptual place as neither a metal nor a
dielectric \cite{Castro09}, the study of graphene's vdW interaction seems worth
of further consideration \cite{Bordag09}.

In this paper I consider the vdW interactions between two graphene layers at
finite temperature ($T$). Graphene, being a critical system at zero $T$, lacks
any characteristic length scale \cite{Castro09}. Temperature provides such
scale, the {\em thermal length}: $\xi_T = \hbar v / k_B T$. We will show that the thermal length controls the vdW
interaction between planes in the form of a crossover. For separations between
the two layers ($z$) smaller than the thermal length,   $z \alt \xi_T$, the
zero-$T$  result ($f$)  for the force \cite{Dobson06}  prevails, $f \propto 1/z^4$.
But for separations larger than the thermal length, $z \agt \xi_T$, the
force  crosses over to a linear-in-$T$ \cite{Lifs56,Landau81}, material
parameters independent, universal regime, $f \propto T/z^3$, that constitutes
the genuine asymptotic large-separation interaction between two graphene sheets
at finite $T$.

The linear-in-$T$ regime will be  shown to reflect the classical nature of
graphene's low lying excitations at finite $T$: plasmons. As shown by Vafek
\cite{Vafek06}, these plasmons are the charge fluctuations of thermally
generated carriers (electrons and holes). Therefore, they are  present only at
finite $T$ and  with energy scale  tied to $T$, so that  long-wavelength
plasmons always behave classically. As such, this thermal regime will be shown
to be present event for the   instantaneous (non-retarded) Coulomb interaction.
This should be contrasted with the usual  linear-in-$T$, thermal limit of the
vdW interactions between {\it any materials} \cite{Lifs56,Landau81} that sets in
for distances larger than  the {\it thermal length of the field},  $\lambda_T =
\hbar c / k_B T$.  The explicit appearance of the light velocity $c$  in this
generic case,  is a manifestation of the classical population of the relevant
electromagnetic modes \cite{Landau81}. But not in graphene, where the existence
of this regime even without retardation $c\rightarrow \infty$, and with the role
of $c$ taken by $v$ in setting the range, shows it to be a  consequence of the
classical dynamics of matter. As a corollary, the inclusion of field's dynamics
(retardation) will be proven to be largely irrelevant both at zero and finite
$T$. This is another aspect where graphene separates from ordinary 
dielectrics and metals, where retardation always matters for large enough
separations. The relevance of the thermal length $(\xi_T)$ (as opposed to $
\lambda_T$)  places graphene in an unique position for the experimental
observation of thermal effects in the vdW interaction. For instance, at room
temperature, the thermal regime should begin to be observable for distances $ z
\agt \xi_T \sim 26 \,\text{nm}$. In contrast, the onset of thermal effects
linked to the classicality of the field (the situation for good metal and
dielectrics, see section \ref{sec:summary}) would require much larger distances
at room temperature:  $z \agt \lambda_T \sim 300 \,\xi_T$.

Concerning the experimental situation, a word of caution is required. Throughout
this paper, graphene is modeled as the usual set of Dirac fermions, known to
provide an excellent description of the low energy physics  \cite{Castro09}
(even up to $\sim 2 \; \text{eV} $ in light absorption experiments \cite{Nair08}).
Nevertheless, the vdW calculation is incomplete for {\it real} graphene, where
more electrons and bands contribute to the vdW force, particularly at short
distances. These additional terms are estimated  from published ab initio
results \cite{Dappe09} in appendix \ref{sec:realgraphene}, and shown not to
compete with the low energy contribution  already for distances of the order of
nanometers.

The paper is organized as follows. In section \ref{sec:formalism}, the formalism
is presented and the results \cite{Dobson06} at zero $T$ recovered. Finite $T$
is considered in section \ref{sec:finite-T},  the force versus scaled distance
($z/\xi_T $) distance is computed numerically and its asymptotic behavior
explained in terms of classical plasmons. Full retardation is included  in
section \ref{sec:retardation}, where its contribution is shown to be
quantitatively irrelevant for graphene's parameters. Section \ref{sec:summary},
summarizes graphene's vdW results, contrasting them with the known behavior of
ideal  dielectrics and metals. Appendix \ref{sec:transverse}  calculates  
graphene's transverse-response contribution to the vdW force, to show that it 
never competes with the the longitudinal response considered in the body of
the paper. Appendix \ref{sec:realgraphene} addresses {\it real} graphene,
estimating the contributions to the vdW force  beyond that of Dirac fermions
considered here.

\section {Formalism. Zero temperature results} \label{sec:formalism}

Let us first present our non-retarded formalism
recovering the zero-$T$ result \cite{Dobson06}. Consider two graphene layers perpendicular to
the z-axis and separated by a distance $z$. Ignoring (for the moment) retardation
effects, the  mutual force per area can be written as:
\begin{equation}\label{force} 
f = \int \frac{d^2 q}{(2 \pi)^2} \: f_c(q,z) 
<\rho_{\bf q}^{(1)}\rho_{-\bf q}^{(2)} > 
,\end{equation}
with Coulomb coupling between density fluctuations
$\rho_{\bf q}^{(1)}$ and $\rho_{-\bf q}^{(2)}$ given by $v_c(q,z) = e^2 \exp(-q
|z|)/(2\epsilon_o q)$ (elementary charge $e$ and vacuum
 permittivity $\epsilon_o$, SI units), and Coulomb force $f_c(q,z)=
-\partial_z v_c(q,z)$.

\begin{figure} % Requires \usepackage{graphicx}
\includegraphics[clip,width=8cm]{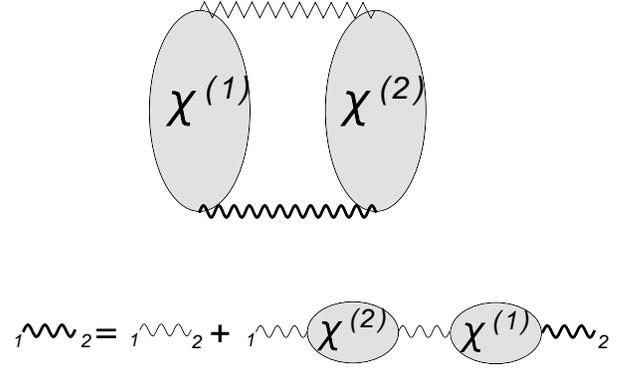} \\ 
 \caption{{\it  Template}
diagram for the computation of the vdW force, as in Eqs. \ref{formula},
\ref{correspondence} and \ref{correspondencetr}. Bubbles: isolated graphene's
response (charge or current). Thin wavy lines: Coulomb interaction (or photon
propagator). Thick wavy lines: multiple-scattering-corrected interaction (or
photon propagator) between graphene's layers as in Eq. \ref{W}. Zig-zag line
represents force ($f_c$): distance derivative of the interaction line (or photon
propagator).} 
\label{feynman} 
\end{figure}

Evaluating the thermal average to all orders in the mutual interaction, we can
write: 
\begin{multline}\label{formula} 
f \! = \!\! -  \frac{1}{(2\pi)^2 }\int \!\! d^2 q 
 \; \frac{f_c(q,z)}{\beta^{-1}} \cdot \\ 
 \cdot \sum_{i \omega_n} \; \chi_{\rho\rho}^{(1)}(q,i \omega_n) \; W_c(q,i
\omega_n,z)  \; \chi_{\rho\rho}^{(2)}(q,i \omega_n) 
,
\end{multline}
with $\beta^{-1}=k_B T$, Matsubara
frequencies $\hbar \omega_n=2 \pi n k_B T $, and the 
multiple-scattering-corrected interaction between planes given by: 
\begin{equation}\label{W}  W_c(q, \omega,z) = \frac{v_c(q,z)} {1 - v_c(q,z) ^2
\chi_{\rho\rho}^{(1)}(q, \omega) \chi_{\rho\rho}^{(2)}(q, \omega)}
,\end{equation} 
 where $\chi_{\rho\rho}^{(1)} =  \chi_{\rho,\rho}^{(2)}) =
\chi_{\rho\rho}$  is the charge-charge  Green's function of an isolated graphene
layer, which can be written as:
\begin{equation}\label{chi} \chi_{\rho\rho}(q,
\omega) =  \frac{\chi_{\rho\rho}^{(0)}(q, \omega)} {1 -
v_c(q,z=0)\chi_{\rho\rho}^{(0)}(q, \omega) }  
,\end{equation}
with $\chi_{\rho\rho}^{(0)}(q,\omega)$ as the polarization of an isolated
graphene ({\em proper} polarization in diagrammatic sense \cite{Fetter}). Before
proceeding to the evaluation of $f$, let us remark that this formula is
entirely equivalent to the (non-retarded version of) Lifshitz
treatment \cite{Landau81}, as can be seen by 
evaluating the field's stress tensor \cite{Pita09} in the presence of the 
(here non-local) material's response. If we knew the exact polarization of a single
graphene (including its crucial $q$-dependence), the only remaining
approximation in (\ref{formula}) (and in Lifshitz's approach) would amount to the
neglect of proper polarization diagrams connecting both planes \cite{Despoja07}:
local field corrections to the dielectric response, safely ignored for large
separations. The diagrammatic description of the vdW calculation is presented
 in Fig. \ref{feynman}.

 Throughout this paper, we will take as the proper polarization the  
 non-interacting value, what amounts to the standard RPA for Eq. (\ref{chi}). It is
given by: 
\begin{equation}\label{chi0} 
 \chi_{ \rho\rho }^{(0)} 
= N \!\!\!\sum_{\sigma, \sigma'=\pm}\! \int \! \! \frac{d^2 k}{(2 \pi)^2}  \:
f^{\sigma \sigma\prime}_{{\mathbf k},{\bf q}} \: \frac{n_f(E_{\bf k}^{\sigma}) -
n_f(E_{{\mathbf k} + {\bf q}}^{\sigma'})} {\hbar \omega - (E_{{\bf k} + {\bf
q}}^{\sigma'} - E_{\bf k}^{\sigma} )}   
,\end{equation} 
with $N\!\!=\!2\!\times\!2$  fermion species, $f^{\sigma \sigma'}_{{\bf k},{\bf q}} =
\frac{1}{2} + \sigma \sigma' \frac{k^2 + {\bf k}\cdot{\bf q}} {2 k |{\bf k} +
{\bf q} |} $, $n_f$ is the Fermi factor and  $E_{\bf k}^{\sigma}  = \sigma \hbar
v k$.   
If we use the zero-$T$ value \cite{Shung86,Gonzalez94} for $\chi^{(0)}$:
\begin{equation}\label{chi0T0} 
 \chi^{(0)}(q,\omega) = -\frac{N}{16 \hbar v}
\frac{q^2}{\sqrt{q ^2 -  \omega^2/v^2}}   
\end{equation} 
in  Eq. (\ref{chi}), with the Matsubara sum becoming an integral 
$i \omega_n \rightarrow i \eta$,  
the resulting expression for $f$ is:

\begin{multline}\label{explicitT0}
 f = - \frac{1}{2 \pi} \int_{0}^{\infty} q^2 dq  \; \frac{\hbar}{\pi}
 \int_{0}^{\infty} d \eta \; \cdot \\
 \cdot \frac{\exp( -2 q |z|)}
 {(1 + \frac{16}{N \alpha} \sqrt{1 + \eta^2/(v q)^2} \;)^2    -\exp( -2 q |z|)}
,\end{multline}
where \mbox{$\alpha=e^2 /(2\epsilon_o \hbar v) $} is a dimensionless
 measure of the
effect of interactions in graphene, with value $\alpha \sim 13.6$.
Expression \ref{explicitT0} can be shown to be entirely equivalent to the
treatment of Dobson {\it et al.} \cite{Dobson06}, leading to the following quantitative  value for
the force per area between graphene layers:
\begin{equation}\label{quantitative}
f = -\frac{A}{z^4} \, , \; \;  A \sim 0.40  \;  \text{eV \AA}
.\end{equation}
Notice that other choices for the proper polarization complying
with the scaling $\chi^{(0)}\propto q f(\omega/vq)$, such as the excitonic
response of ref.~\onlinecite{Farid08}, would produce the same
$z^{-4}$ power law, although with different prefactor.  

\section {Finite temperature results} \label{sec:finite-T}

\begin{figure} % Requires \usepackage{graphicx}
\includegraphics[clip,width=8cm]{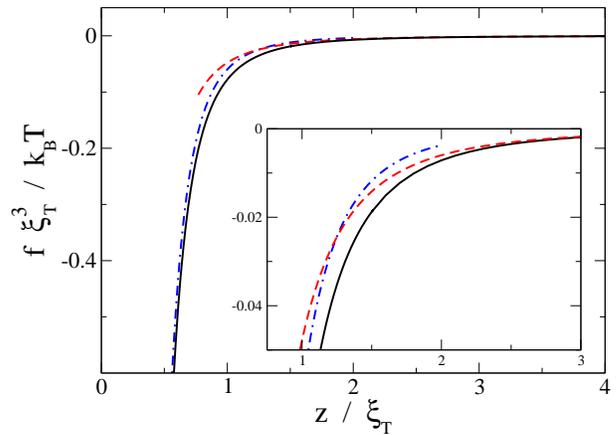} \\ 
\caption{van der Waals (scaled)
force per area $\tilde{f}$ between graphene planes versus distance $z$ in units of the
thermal length $\xi_T$. Continuous line: numerical result. Dashed line: large
distance $(z \gg \xi_T)$ limit, Eq. (\ref{central}). Dashed-dotted line:
zero-temperature limit $(z \ll \xi_T)$, (Eq. \ref{quantitative}). Inset: enlarged
view of the crossover region $(z \sim \xi_T)$.}

\label{vdw}
\end{figure}

Let's consider now a finite temperature still for non-retarded interactions. 
Although no simple analytical 
expression is
known for $\chi^{(0)}$ at finite $T$ 
(see, though, ref.~\onlinecite{Vafekthesis}), its scaling behavior \cite{Vafek06} is best described
measuring lengths in units of $\xi_T = \hbar v / k_B T$, and energy in terms of
$k_B T$. Indeed, for the force calculation,  matter and field appear in the
dimensionless combination:
\begin{equation}\label{dimensionless}
  v_c(q,z) \; \chi^{(0)}(q,\omega) = 
  \alpha \; \exp( -q |z|) \; \tilde{\chi}(q \: \xi_T,\hbar \omega /
  k_B T )
,\end{equation}
where
$\tilde{\chi}(q\xi_T,\hbar \omega / k_B T ) $ is a dimensionless function. It
is  clear that the force, Eq. (\ref{formula}), will depend on
distance and temperature only through the combination $z / \xi_T$, with the
following scaling form:  
\begin{equation}\label{fdimensioless}
f(z,T) = \frac{k_B T}{\xi_T^3} \tilde{f}(z / \xi_T)
\end{equation}
Therefore, knowledge of the dimensionless function $\tilde{f}(z / \xi_T)$
provides all information for the vdW interaction at finite $T$ and arbitrary
distances in the scaling regime. We have evaluated numerically the force (Eqs.
(\ref{chi0}), (\ref{chi}) and (\ref{formula})) with results  plotted in Fig.
\ref{vdw}. As expected,  graphene's thermal length $\xi_T$ marks a crossover
between two regimes: the zero-$T$  limit (\ref{quantitative}) for $ z/\xi_T \ll 1$ 
 previously analyzed \cite{Dobson06}, and the genuine large-distance
regime at finite temperature for $ z/\xi_T \gg 1$, that we now consider.

As discussed by Vafek \cite{Vafek06}, the most important feature of the charge-charge response
 (\ref{chi})  at finite $T$, is the emergence of plasmons. These appear as the
zeros of the denominator of Eq. (\ref{chi})  
 and, in the long wavelength limit, the
plasmon frequency is given by \cite{Vafek06,wpret}:  
\begin{equation}\label{wp}  
\frac{\hbar \omega_p(q)}{k_B T } = \sqrt{\frac{\ln 2}{2 \pi}\: N \:\alpha 
 \: q \:\xi_T}, \; \;  q \: \xi_T \alt 1 
\end{equation} 
Although plasmons possess an imaginary part $\gamma(q)$, meaning that they 
decay into the  electron-hole continuum \cite{Vafek06}, they  become very long-lived
excitations for long wavelengths: $ \gamma(q) / \omega_p(q) \rightarrow
0$  for  $q \,\xi_T \rightarrow 0 $.

 Being the density fluctuations of  thermally excited carriers, plasmons owe
their existence and  energy scale  
%(Eq. \ref{wp})  
to temperature. In this
respect, they differ from plasmons of ordinary 2d metals, already present 
and contributing \cite{Sernelius98,Dobson06}
 to the vdW force at zero $T$.
Furthermore,
the spectral power of the charge-charge response for  $\hbar v q < \hbar\omega
\alt k_B T $ and $q \: \xi_T  \ll 1 $,  is dominated by the plasmon mode.
Therefore, a single-pole approach for the response suffices for the plasmon
contribution to $ \chi $ at finite $T$, with the explicit form: 
\begin{equation}\label{chiplasmon} 
\chi(q,\omega) = \frac{1}{v_c(q,z=0)}
\frac{\omega_p(q)^2} {\omega^2 -\omega_p(q)^2} 
\end{equation}  
This response is valid for $q \: \xi_T \ll 1 $ and 
$\hbar\omega \alt k_B T $. Its use in Eq. (\ref{formula})
provides the wanted  large-distance  
% ($z /\xi_T \gg 1 $)
behavior of the vdW
force at finite $T$. The evaluation is best performed trading Matsubara sums
for real frequency integration in  Eq.~(\ref{formula}):
\begin{multline}\label{explicitT}
 f = - \frac{1}{2 \pi} \int_{0}^{\infty} q^2 dq  \; \frac{\hbar}{\pi}
 \int_{0}^{\infty} d \omega \; \coth(\beta\hbar\omega/2) \cdot \\
 \cdot \Im  \; \frac{\exp( -2 q |z|)}
  {(\frac{\omega^2}{\omega_p(q)^2} -1 )^2    -\exp( -2 q |z|)}, 
\end{multline}
leading to the following
central result:
\begin{equation}\label{central}
f = - \frac{\zeta(3)}{8 \pi} \frac{k_B T}{z^3}, \;\; z\gg\xi_T
,\end{equation}
with the Riemann's zeta function, $\zeta(3) = 1.2020... \: $.
Eq.~(\ref{central}) is the large-distance asymptotic behavior for the force per
area of two graphene sheets at finite temperature. 
The numerical solution does indeed merge with this analytical limit for
$z\gg\xi_T$, as seen in Fig. \ref{vdw}.

The result of Eq. (\ref{central}) is truly remarkable:   all material and
electrical parameters have disappeared, leaving  the temperature as the only
surviving  energy scale. As remarked in the introduction, an  identical formula
describes the force between two metallic plates at finite temperature, for
distances larger than the thermal length of the electromagnetic field \cite{Landau81}, $ z  \gg
\lambda_T = \hbar c /k_B T$. This is  the limit where the thermal
population of the relevant electromagnetic modes becomes classical. But, in
spite of  the similarity,  we cannot make an obvious connection with our
result:  our treatment has been obtained for the {\em instantaneous},
non-retarded Coulomb interaction, therefore there is no field dynamics, no field
modes, and the issue of classicality for the field is out of place. Setting
$c=\infty$ in $\lambda_T $ renders meaningless the would-be range for that
classical limit. Yet, our regime of Eq. (\ref{central}) for the instantaneous
interaction appears for  $z \agt \xi_T = \hbar v /k_B T$.

Nevertheless, the fact that $v$ takes the role of $c$ in setting the range for
our non-retarded calculation prompts for  the existence of a 
classical interpretation, but now for the only dynamical entity so far considered:
matter.   Plasmons, by the very fact that their existence and scale are tied to
temperature, behave classically at long wavelengths:
\begin{equation}\label{classicality}
\frac{\hbar \omega_p(q)}{k_B T} \rightarrow 0, \; \;   q \: \xi_T \ll 1
,\end{equation}
and this  suggests that there must be more transparent ways of getting such a
simple result as Eq. (\ref{central}). As  reassurance that our reasoning
is well founded, we will now recover Eq. (\ref{central}) invoking  only
elementary classical concepts. Let's consider graphene's  charge fluctuations
as classical objects at temperature $T$. The classical limit means that we can
ignore kinetic energies and rely only on the potential (electrostatic)
energy to account for the thermal population of these fluctuations. This 
electrostatic energy is: 
\begin{eqnarray}\label{Uel}
U_{el} \!\!& = & \! \!\sum_{\bf q} v_c(q,z) \rho_{\bf q}^{(1)}\rho^{(2)}_{-\bf q} + 
\frac{1}{2}  v_c(q,0) (\rho_{\bf q}^{(1)}\rho^{(1)}_{-\bf q} + 
 \rho_{\bf q}^{(2)}\rho^{(2)}_{-\bf q})  \nonumber \\ 
& = & \sum_{\bf q} \sum_{\sigma = \pm}
 \frac{1}{2}  v_{\sigma}(q,z) \rho_{\bf q}^{(\sigma)}\rho^{(\sigma)}_{-\bf q}
\end{eqnarray} 
where we have diagonalized the quadratic form with the normal modes:  
$\rho_{\bf q}^{(\pm)} = (1/\sqrt{2})(\rho_{\bf q}^{(1)} \pm \rho_{\bf
q}^{(2)})$, with $v_{\pm}(q,z) = v_c(q,0) \pm v_c(q,z)$. The 
equipartition theorem allows us to write the thermal population of    modes as 
$ <\rho_{\bf q}^{(\pm)}\rho^{(\pm)}_{-\bf q}> = k_B T /v_{\pm}(q,z) $. 
Expressing $\rho_{\bf q}^{(1,2)}$ in terms of $\rho_{\bf q}^{(\pm)}$, the
thermal average of  Eq. (\ref{force}) can be obtained with the result of 
Eq. (\ref{central}). This fully supports our interpretation that 
it is the classical
population of thermal plasmons what leads to the vdW force. 

 Preparing for the discussion of retardation in the next section, it is worth
 noticing that the expression of Eq. \ref{central}  amounts to selecting the  $
 i \omega_n =0$ term  in the frequency sum of Eq. \ref{formula}. Indeed, this is
 expected for a classical limit, but notice that the reason for such behavior is
 entirely due to the matter response $\chi$ and, again,  has nothing to do with
 the interacting field which, considered so far instantaneous, therefore
 shows no frequency dependence.

\section {Retardation} \label{sec:retardation}

 Now we address the issue of the electromagnetic field
dynamics, to show that the inclusion of retardation hardly affects the previous
results. Following ref.~\onlinecite{Landau81}, retardation effects are best handled in a
 gauge  where
only the vector potential ${\bf A}$ exists. The coupling matter-field is of the
form $\propto {\bf j}\cdot{\bf A}$, where the current lays in graphene's
planes, and can be decomposed into (in-plane) longitudinal and transverse components that
are not mixed by the photon field. Let's consider the longitudinal current
responsible for charge fluctuations. It is straightforward to show that
retardation can be included in the previous formalism  with the following
correspondences in Eqs. (\ref{formula}) and (\ref{W}):
\begin{eqnarray}\label{correspondence}
\chi_{\rho\rho}(q,\omega) &\rightarrow& \chi_{j_l j_l}(q, \omega) \nonumber \\ 
v_c(q,z) &\rightarrow& {\cal D}_l(q, \omega,z) \\
f_c(q,z) &\rightarrow& -\partial_z {\cal D}_l(q, \omega,z) \nonumber
\end{eqnarray}
where ${\cal D}_l(q,\omega,z)$ is the (part of the) photon propagator that
couples to in-plane longitudinal currents, with expression:
\begin{equation}\label{photon}
{\cal D}_l(q,\omega,z) = \frac{e^2 q'  \exp(-q' |z|)}{2 \: \epsilon_o \:
\omega^2}
,\end{equation}
and $ q' = \sqrt{ q^2 - \omega ^2 / c^2}$.  $\chi_{j_l j_l}$ is the longitudinal
current-current response, related by particle conservation to the charge-charge
response by  $ q^2 \chi_{j_l j_l}(q,\omega) = \omega^2 \chi_{\rho\rho}(q,\omega)$. 
It can be checked that setting the light velocity $c \rightarrow \infty$ in the
above expressions, the non-retarded expression for the force is recovered.
Again, as in the non-retarded case, this formalism for the force can be shown
to be exactly equivalent to Lifshitz's when applied to the longitudinal
response.

Let's consider the results for zero temperature first. Carrying out the
prescription of Eq. (\ref{correspondence}) and trading $q$ for $q'$, the
following expression is obtained for the vdW force due to longitudinal
currents, including retardation: 
\begin{multline}\label{explicitretard}
 f = - \frac{1}{2 \pi} \int_{0}^{\infty} q'^2 dq'  \; \frac{\hbar}{\pi}
 \int_{0}^{c q'} d \eta \; \cdot \\
 \cdot \frac{\exp( -2 q' |z|)}
 {(1 + \frac{16}{N \alpha} \sqrt{1 + \eta^2/(v_{eff} q')^2} \;)^2    -\exp( -2 q' |z|)}
\end{multline}
where $v_{eff}^{-2} = v^{-2} - c^{-2} $. 
Comparing Eqs. \ref{explicitretard} and \ref{explicitT0}, one sees that 
only two formal changes appear with respect
to  the zero-$T$, non-retarded calculation. 
 First, there is a renormalization of graphene's velocity in the
square root of Eq. (\ref{chi0T0}), 
%$v \rightarrow v_{eff} = v / \sqrt{1 + v^2 /c^2} $,
$v \rightarrow  v / \sqrt{1 - v^2 /c^2} $, 
quantitatively irrelevant. Second, the integration over imaginary
frequency acquires an upper limit, whose physical interpretation corresponds to the
removal of the electromagnetic field modes that, for each space scale, are
slower than matter. This is the dominant effect of retardation  but, the
integrand decaying as  $\sim \eta^{-2}$, it amounts to a meager  $ v/c
\sim (300)^{-1}$  fractional reduction of the prefactor $A$ in Eq.
(\ref{quantitative}) without  altering the power law.

\begin{figure} % Requires \usepackage{graphicx}
\includegraphics[clip,width=8cm]{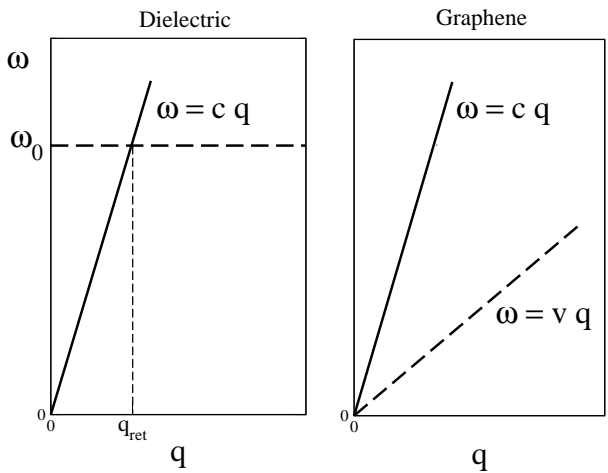} \\ 
\caption{Left panel: light-cone ($\omega = c q$) versus dynamics of a dielectric
 of typical frequency $\omega_0$, illustrating the importance of retardation for
 distances beyond $z_{ret}\sim q^{-1}_{ret}$. Right panel: light-cone versus
 graphene's typical dynamics ($ \omega = v q$, not to scale) illustrating
  the absence of a 
 characteristic distance for retardation.}

\label{retardation}
\end{figure}

The irrelevance of retardation in graphene at zero $T$  contrasts  with the
situation for a regular dielectric, where retardation always matters beyond some
distance $z_{ret} \sim q_{ret}^{-1}$, with $q_{ret} \sim \omega_o / c $, where
$\hbar \omega_o$ is a typical energy scale (say the gap). 
This is schematically depicted in the left panel of Fig. \ref{retardation}. 
In graphene (right panel of Fig. \ref{retardation}), on the
contrary, both matter and field are scale-invariant (critical) systems (with
dynamical critical exponent 1), this implies that the ratio ($c/v\sim 300$) of
their relative dynamics remains the same at every length scale (separation
between planes).  Therefore, the  irrelevance of retardation effects in graphene at 
zero $T$ is both qualitative and quantitative. Qualitative because, at least within
our RPA treatment, the power law for the vdW force of Eq. (\ref{quantitative})
would remain the same for arbitrary values of graphene's velocity $v$, although
with a changed prefactor. For graphene, this irrelevance is also
quantitative, because the prefactor barely changes: $ \delta A / A \sim 1/300
$.

Now we show that  retardation at finite $T$ also lets unaffected Eq. (\ref{central}) as
the correct large distance behavior.  The spatial dependence of the photon
propagator (\ref{photon}) makes short-ranged the contribution from Matsubara
frequencies other than  $n=0$. This effect begins to matter  for  distances  $ z
\agt \lambda_T = \hbar c / k_B T$, and is present in any material as it 
corresponds to the above mentioned classical limit of the thermal population of
electromagnetic modes. But for graphene, restricting to $n=0$ adds nothing to
the  non-retarded result for Eq. (\ref{central}). Indeed, such result is
equivalent to selecting $n=0$ in the matter response, although in that case this
restriction was forced upon us by the classical behavior of matter's plasmons
while the field remained instantaneous. In other words, for the vdW interaction
in graphene, there is no difference between {\it classical matter +
instantaneous field} and {\it classical field}.  We have computed the force 
with the numerically evaluated, finite-$T$ response of Eq. (\ref{chi0}) and the
retarded interaction,  with results  that would be
hardly distinguishable from the non-retarded curve shown in Fig. (\ref{vdw}). 

A
further contribution to the vdW force exists from the coupling of the field to 
transverse currents. In fact, for good dielectrics and metals and in the region
where retardation is important, transverse and longitudinal current fluctuations
contribute similarly to the vdW force, as discussed in section \ref{sec:summary}.
But not for graphene: in appendix \ref{sec:transverse} this transverse
contribution is calculated and shown
 to be, at best, of the order of $v/c$ times smaller 
than the longitudinal part, a result consistent with the
absence of a retarded regime for the longitudinal part.

\section{Summary: G\lowercase{raphene vs} D\lowercase{ielectric and}
 M\lowercase{etal} } \label{sec:summary}

Here we summarize graphene's results for the vdW interaction and, to gain a
better perspective,  compare them with the standard prototypes of metals and
dielectrics. The basic results for graphene, Eq. \ref{quantitative} and Eq.
\ref{central}, are collected here: 

\begin{equation} \label{graphene}
\begin{array}{lr} 
   f \propto -\hbar \, v \; z^{-4}, &  z \ll \xi_T  
 \\ [1pt]
   f = - \frac{\zeta(3)}{8 \pi} \frac{k_B T}{z^3}, & z\gg\xi_T
 \end{array}
\end{equation} 
We have seen that  a single length, graphene's thermal length:  $\xi_T = \hbar
v / k_B T $, controls the vdW force, which exhibits a crossover from the
zero-$T$ results of  Dobson {\it et al.} \cite{Dobson06} (linked to the linear
dispersion of Dirac fermions) to a finite-$T$,  universal regime. The latter 
has been understood as  arising form the existence of {\it classical} plasmons:
 charge fluctuations of thermally generated carriers whose existence and
energy scale are tied to temperature. These results, originally obtained for the
longitudinal (charge) response with instantaneous Coulomb coupling, have been
shown to survive virtually unaffected when retardation is included. In
addition, the transverse contribution is shown in the appendix 
\ref{sec:transverse} to never compete
with the longitudinal one. Finally, we have emphasized that the temperature
dependence of the vdW interaction in graphene reflects basically a property of
matter, as opposed to the corresponding thermal regime of the archetypal metals
and dielectrics that are described in the following subsections.

\begin{figure} % Requires \usepackage{graphicx}
\includegraphics[clip,width=8cm]{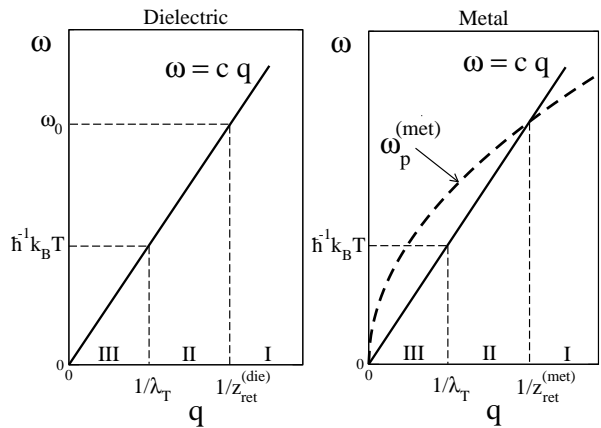} \\  
\caption{Left
panel: light-cone ($\omega = c q$) versus dynamics of a dielectric of typical
frequency $\omega_0$, illustrating the three vdw regimes encountered with
increasing distance at finite temperature (Eq. \ref{dielectric}).  Right panel:
As in left panel but for a typical 2d metal, with dynamics characterized by
plasmons (Eq. \ref{metal}).}

\label{fig:comparison}
\end{figure}

\subsection{Graphene vs Dielectric} 

To make a meaningful comparison, we consider a two-dimensional insulator, with
 characteristic frequency (gap)  and length scale of atomic dimensions:  $ \hbar
 \omega_0 \sim \text{eV} $ and $ a_0 \sim \text{\AA}$. We assume {\it low}
 temperatures, so that the matter response is well approximated by the zero-$T$
 response (room temperature is low temperature for an eV gap). In this
 situation, three regimes \cite{Landau81} can be considered in the vdW force as a function of
 separation, corresponding to the $q$ regions schematically shown in  the left
 panel of Fig. \ref{fig:comparison}:
\begin{equation} \label{dielectric}
\begin{array}{lcr} 
 I: & f \propto - \hbar \, \omega_0 \,a_0^2 \; z^{-5}, &  z \alt z^{(die)}_{ret}  
 \\ [1pt]
 II: & f \propto - \hbar\, c\, a_0^2 \; z^{-6}, &  z^{(die)}_{ret} \alt z 
 \alt  \lambda_T  
 \\ [1pt]
 III: & f \propto - k_B T \,a_0^2 \; z^{-5}, &  z \agt \lambda_T 
\end{array}
\end{equation} 
where the characteristic distance for the onset of retarded effects is 
$z^{(die)}_{ret} = c/\omega_0$, with the field's thermal length as before: 
$ \lambda_T = \hbar c / k_B T$. 
Notice that Fig. \ref{fig:comparison} and  Eq. \ref{dielectric} are schematic: transitions
between regimes are in the form of crossovers, and the very existence of a
regime (particularly II), and its associated power law, assumes well separated
values of  $z^{(die)}_{ret}$ and $\lambda_T$.

 Regime I corresponds to the {\it sum of $R^{-6}$} contributions expected from
a dielectric, where  longitudinal (charge) fluctuations, instantaneously
coupled, dominate the force. In region II, matter is quicker than field, and
retardation matters, implying and additional power with distance. Here both
longitudinal and transverse currents contribute similarly to the vdW force.
Finally, a thermal regime (III)  appears for distances  beyond $\lambda_T$, at
which  the short-ranged nature of the coupling due the electromagnetic modes
with finite Matsubara frequencies leaves the mode $i \omega_n = 0$ as the sole
contribution to the vdW force. Notice that the $T$ dependence of the vdW force
in this thermal regime is not due any $T$-dependence in the matter response,
which remains that of zero-$T$ for the {\it low} temperatures considered, as
explained before. 

The comparison of Eq. (\ref{graphene}) (and accompanying discussion) and Eq.
(\ref{dielectric})  clearly exhibits  that, for graphene, it is a difference in
the dominant matter response what changes  the vdW force at finite $T$ for
distances above  the  ($T$-dependent) thermal length, $\xi_T$. For shorter
distances, the zero-$T$ response dominates while at larger distances, the
coupling of classical charge fluctuations of thermally excited carriers
dominates. Neither retardation nor transverse  currents 
(see appendix \ref{sec:transverse}) do matter
quantitatively, what explains the absence in graphene  of a retarded regime like
region II of a dielectric.

\subsection{Graphene vs Metal} 

 We consider a two-dimensional gas of electrons, treated in the RPA
approximation and characterized by a Fermi wavevector typical of a good metal $
k_f \sim \mbox{\AA}^{-1} $, with Fermi velocity of the  order of
 graphene's  $ v \sim c / 300$ (certainly  of the
order of a typical metal). 
Interactions are then characterized by the same dimensionless number
 as in graphene:
$\alpha=e^2 /(2\epsilon_o \hbar v) \sim 13.6 $, and just for ease, we take
$N = 4 $ species of fermions, as in graphene. As for the dielectric, we assume
a sufficiently low temperature such that the matter response is basically that
of zero-$T$ (a good degenerate metal). Again, three separation regimes can be
considered \cite{Landau81,Sernelius98}, with corresponding $q$ regions schematically depicted in the right
panel of Fig. \ref{fig:comparison}:
\begin{equation} \label{metal}
\begin{array}{lcr} 
 I: & f \propto - \hbar \, v \, k_f^{1/2}\; z^{-7/2}, &  z \alt z^{(met)}_{ret}  
 \\ [2pt]
 II: \; (Casimir) & f = - \frac{\pi^2}{240} \frac{\hbar\, c}{z^4}, &  z^{(met)}_{ret} \alt z 
 \alt  \lambda_T  
 \\ [3pt]
 III: & f = - \frac{\zeta(3)}{8 \pi} \frac{k_B T}{z^3}, & z \agt \lambda_T
\end{array}
\end{equation} 
where the characteristic distance for the onset of retarded effects in the
metal is 
$z^{(met)}_{ret} \sim  (\frac{N \alpha}{4 \pi} \frac{v^2}{c^2} k_f)^{-1}$,
 and the field's thermal length as before: 
$ \lambda_T = \hbar c / k_B T$. 
Again, the regimes (mainly II) of Eq. \ref{metal} are meaningful only for well separated values
of  $z^{(met)}_{ret}$ and $\lambda_T$.

  Regime I corresponds to the instantaneous coupling of plasmons, as discussed
in refs.~\onlinecite{Sernelius98} and \onlinecite{Dobson06}. It is very important to realize that
plasmons, with frequency  $\omega^{(met)}_p(q) = \sqrt{\frac{N \alpha}{4 \pi}
v^2 k_f q}$,  are an intrinsic feature of  metals at zero $T$. At the
low-$T$  considered here and for wavevectors in region I (see Fig.
\ref{fig:comparison}), they are
certainly quantum objects, $\hbar \omega^{(met)}_p  \gg k_B T  $. This is in
contrast with the situation for graphene, where plasmons only exist at
finite $T$  and with frequency tied to temperature (see Eq. \ref{wp}), therefore
they are always classical:  $\hbar \omega^{(graphene)}_p  \ll k_B T  $. This
explains the  absence of a $z^{-7/2} $ vdW regime in graphene, whose presence
would require free carriers at zero-$T$, certainly not the case in graphene, where
this zero-$T$ regime is replaced by Dobson {\it et al.} \cite{Dobson06} $1/z^4$
force.

 Regime II in Eq. \ref{metal} is the well known Casimir result. Here the field
is quicker than matter (plasmons) and a  fully retarded calculation is needed
(see note \onlinecite{wpretmet}),
with  longitudinal and transverse currents in the 2d metal  contributing
equally to the Casimir universal result. This regime has no equivalent in
graphene. Again, the reason is the classical behavior of charge fluctuations in
graphene. The existence of a Casimir regime would require simultaneously  $ 
\omega^{(graphene)}_p(q) \gg c q$  and  $ \hbar \omega^{(graphene)}_p  \gg k_B
T $, something impossible for graphene's temperature-tied plasmons (see Eq.
\ref{wp}). The quantitative irrelevance of the transverse response for the vdW
force in graphene, explained in the appendix \ref{sec:transverse},
 is also consistent with the
absence of this Casimir regime (transverse currents are responsible of half the
Casimir result in the 2d metal).  

Finally, thermal effects do appear in regime III, where the result for metals is
quantitatively identical to that of graphene, Eq. \ref{central}. But there are
profound differences in the reasons that lead to the same expression in both
systems. In the 2d metal, the matter response remains that of zero-$T$, but
temperature forces to select the $ i \omega_n = 0 $ term  as the only
electromagnetic mode that is not short-ranged. Therefore, although the result
itself, Eq. (\ref{metal}, III), cannot exhibit the dynamics of the field, the
explicit appearance of the light velocity $c$ in its range of applicability ($z
\agt \lambda_T=\hbar c /k_B T  $) does reveal its origin. In contrast,
graphene's finite-$T$ behavior Eq. \ref{central} (obtained for the {\it
instantaneous} Coulomb coupling),  has been shown to arise from the classical
plasmonic response of matter, as explained before. Therefore the range of
applicability of the thermal regime in graphene $( z \agt \xi_T=\hbar v
/k_B T ) $ does not contain the light velocity (infinite for non-retarded
calculation), but a matter property:  Dirac fermion's velocity $(v)$.

  Let us close mentioning that there has recently been much interest  in the
issue of finite-$T$  vdW interactions in {\em poor} metals and its relation to
non locality in the metal's response \cite{Sernelius06,Pita08,Dalvit08,Sveto08}.
Graphene may well provide a natural ground for these concerns as a system
exhibiting both a  zero-$T$,  dispersive-response, result and a classical 
linear-in-$T$ regime, but at much shorter distances than would otherwise be
required for the classicality of the electromagnetic field: at room temperature,
$ \xi_T \sim 26 \: \text{nm}$ for graphene versus $ \lambda_T \sim 300 \;\xi_T$ for 
typical metals and dielectrics.

\begin{acknowledgments}
I am thankful to L. Brey, F. Guinea, P. L\'opez-Sancho, M.A.H. Vozmediano, and
F. Yndur\'ain for useful discussions. Financial support from Spain's MEC project
FIS2005-05478-C02-01 is acknowledged.
\end{acknowledgments}

\appendix 

\section{Transverse currents contribution} \label{sec:transverse}

Here we present the calculation of the force due to the (in-plane) transverse
currents, to show that it never competes with the longitudinal part. As in
section \ref{sec:retardation}, the force per area can be obtained with the following
correspondence in Eqs. (\ref{formula}) and (\ref{W}):

\begin{eqnarray}\label{correspondencetr}
\chi_{\rho\rho}(q,\omega) &\rightarrow& \chi_{j_t j_t}(q, \omega) \nonumber \\ 
v_c(q,z) &\rightarrow& {\cal D}_t(q, \omega,z) \\
f_c(q,z) &\rightarrow& -\partial_z {\cal D}_t(q, \omega,z) \nonumber
\end{eqnarray}
where ${\cal D}_t(q,\omega,z)$ is the (part of the) photon propagator that
couples to in-plane transverse currents, with expression: 
\begin{equation}\label{photont} 
{\cal D}_t(q,\omega,z) = - \frac{e^2  
\exp(-q' |z|)}{2 \: \epsilon_o \: q' c^2} ,
\end{equation} 
% %\end{document} %
and $ q' = \sqrt{ q^2 - \omega ^2 / c^2}$.  $\chi_{j_t j_t}$ is the transverse
current-current response of an isolated graphene layer, with RPA expression
given by:
\begin{equation}\label{chitr} 
\chi_{j_t j_t}(q,\omega) =  
\frac{\chi_{j_t j_t}^{(0)}(q, \omega)} {1 -
{\cal D}_t(q,\omega,z=0)\chi_{j_t j_t}^{(0)}(q, \omega) }  
,\end{equation}
where the  zero-$T$,  non-interacting response is given by
\begin{equation}\label{chi0T0tr} 
 \chi_{j_t j_t}^{(0)} = \frac{N v}{16 \hbar }
{\sqrt{q ^2 -  \omega^2/v^2}}.   
\end{equation} 

For graphene parameters, $(c\sim 300 \; v)$, it is straightforward to see that
the the RPA treatment is unnecessary for the response, therefore we take $
\chi_{j_t j_t} \sim\chi_{j_t j_t}^{(0)}$. Trading $q$ for $q'$, with the Matsubara sum becoming an integral 
$i \omega_n \rightarrow i \eta$,  
the resulting expression for the force $f_t$ is: 
\begin{multline}\label{explicitft}
 f_t = - \frac{1}{2 \pi} \int_{0}^{\infty} q'^2 dq'  \; \exp( -2 q' |z|)
  \; \frac{\hbar}{\pi}
 \int_{0}^{c q'} d \eta \;  \cdot \\
 \cdot 
 \left( \frac{N \alpha}{16} \frac{v^2}{c^2} \right) ^2 
 \left( 1 + \frac{\eta^2}{v_{eff}^2 q'^2}\right)
\end{multline}
where, as in the main text, $v_{eff}^{-2} = v^{-2} - c^{-2} $, and 
\mbox{$\alpha=e^2 /(2\epsilon_o \hbar  v) $}. 
Ignoring  the meager renormalization implied by $v_{eff}$, and to lowest order
in $v/c$, the following closed expression for the force is obtained: 

\begin{equation}\label{forcet} 
f_t = - \frac{(N \alpha)^2}{2^{12} \, \pi^2} \frac{v}{c} \frac{\hbar \, v}{z^4}
\end{equation} 

Although the dependence with distance
 of  Eq.~(\ref{forcet}) is the same as that of the longitudinal
contribution, Eq. \ref{quantitative},  the salient feature is the ratio $v/c$
in the coefficient. Indeed,
explicit comparison of Eq.(\ref{forcet}) and  Eq. \ref{quantitative} leads to:  

\begin{equation}\label{comparison}  
f_t \sim 1.23 \;\frac{v}{c} \; f \sim 4.1
\times 10^{-3} f  
\end{equation}  % 

Therefore, the neglect of the transverse contribution is fully justified. Even
more, our result (Eq. \ref{comparison}) can be seen as a {\it bound} for the
transverse force over an extended range of distances. The reason is that the
transverse contribution here calculated uses the scaling form for graphene's 
response, Eq. \ref{chi0T0tr}, but this response grows unbounded with
(imaginary) frequency. Real graphene lives in a lattice and, therefore, we
expect this response to saturate beyond a cut-off of the order of  $\hbar
\omega_0 \sim \text{eV}$ (the standard diamagnetic local limit). This implies that our
zero-$T$ calculation (\ref{forcet}) should apply for distances 
$ z \agt  c / \omega_0$, while for shorter distances, the saturation of matter response must
imply even lower (cut-off dependent) forces from the transverse response.

Finally, we consider the transverse contribution at finite $T$. It suffices to
notice that the zero-$T$ result (\ref{explicitft}) is dominated by the matter
response at (imaginary) frequencies close to the light cone: $\eta \sim c \,
q$. At such high frequencies, temperature hardly affects the matter response,
and the zero-$T$ transverse contribution holds for distances up to  $z \agt
\lambda_T $ $(\lambda_T = \hbar c / k_B T)$. Beyond such distances, as usual,
only the  $ i \omega_n = 0$ term survives in the frequency sum (Eq.
\ref{formula}) as the long-range interaction between graphene's layers. Then,
using the long-wavelength, finite-$T$ result for the static transverse response:

\begin{equation}\label{chi0T} 
 \chi_{j_t j_t}^{(0)}(q, i\omega_n = 0) = 
 \frac{N}{24 \pi} \frac{(v \; q)^2}{k_B T}, \; \; q \, \xi_T \ll 1,
\end{equation} 
we end up with the following expression for the large-distance,
transverse contribution to the force per area  between graphene's layers at finite
temperatures:

\begin{equation}\label{forcetT} 
f_t = -\frac{(N \,\alpha \, \xi_T)^2  \, k_B T}{1356 \, \pi^3} \,\frac{v^4}{c^4}
\, \frac{1}{z^5}, \; \;
z \agt \lambda_T 
\end{equation} 

This force replaces the zero-$T$ expression (\ref{forcet}) for distances 
$z \agt \lambda_T$. Quantitatively, it
is utterly small when compared with the corresponding finite-$T$
longitudinal contribution (Eq. \ref{central}). This completes our justification
for the neglect of the transverse contribution in the body of the paper, both
at zero and finite temperatures.

\section{Real graphene versus Dirac fermions} \label{sec:realgraphene}

The success of the Dirac fermion approach to explain the experimentally observed
electronic properties of graphene leaves little doubt about its correctness for
describing the {\em low} energy physics \cite{Castro09}, including room
temperature and above. The long distance behavior of the vdW interaction cannot
be an exception: qualitative features such as the asymptotic power law of
Eq.~\ref{quantitative} are linked to the linear dispersion of Dirac fermions 
and its associated gapless excitations, as first shown by Dobson {\it et al.}
\cite{Dobson06}. On the contrary, {\it sum of $R^{-6}$} treatments would lead to
an {\em incorrect} $1/z^5$ {\em dielectric} law for the force. 

Nevertheless, the presence of further electrons (and states) besides those
of Dirac cones makes our calculation quantitatively incomplete as far as
{\em real} graphene is concerned. 
These
electron-hole  excitations being gapped,  their contribution to the vdW
interactions is not expected to change the obtained asymptotic behavior.
Nevertheless, estimating the size of this  neglected
contribution  is important to assess the range of validity of our results.

Explicit calculation of these contributions is well beyond the scope of this
work, but a fair estimation of its size can be obtained from published
results\cite{Dappe09}. Such an estimation can be restricted to zero $T$ because:
i) the missing contribution corresponds to gapped excitations above $\sim 2 \,
\text{eV}$, hardly affected  at any reasonable finite $T$, ii) our results at
finite $T$, Fig. (\ref{vdw}), are always greater than those at zero $T$.
Therefore,  if the missing contributions do not compete with the asymptotic
regime at zero $T$, they will not  at finite $T$. 

In ref.~\onlinecite{Dappe09}, an {\it ab initio} calculation of the vdW interaction
in graphene-like systems is performed. Not intended for large distances, the
low energy physics in ref.~\onlinecite{Dappe09} is only approximated, leading to an
effective  long-distance interaction between carbon atoms of the traditional
dielectric form: $1/R^{-6}$. Nevertheless, ref.~\onlinecite{Dappe09}  includes
electrons beyond the lowest {\it sp} band, and high energy excitations (up to
tens on eV)\cite{Dappe09}, providing a quantitative estimate of the   missing
contributions. The attractive energy between two atoms in different graphene
sheets separated by $R$ is obtained there \cite{Dappe09}  as  $C_6 / R^6$, with   $C_6 \sim 13.8 \;
\text{eV \AA}^6$,   and  around $50$  percent of this contribution is reported
as coming from excitations beyond the lowest  {\it sp} band \cite{Dappe09}.
From these numbers, one obtains an estimate of $ f = -D_6/z^5$, with $D_6 \sim
6.3 \; \text{eV \AA}^2$, for the force per area coming from the neglected terms.
This contribution becomes comparable to that of Eq. \ref{quantitative} only at 
separations between graphene layers of the order of $1.5 \,\text{nm}$, a rather
short distance for a vdW scenario, just a few times graphite interlayer
distance. Beyond this distance, the different power law makes the asymptotic
contribution of Eq. \ref{quantitative} dominant (see also
ref.~\onlinecite{Sabio08}). For instance, at a distance $z \sim 27 \,\text{nm}$,
corresponding to graphene's thermal length a room $T$, the asymptotic
contribution here obtained (Fig. \ref{vdw})  is around $20$ times the estimate
from the ignored terms. At a separation of  $ z \sim 100 \, \text{nm} $ (graphene's
thermal length at  liquid-nitrogen $T$), the force calculated in this work,  
would surpass the missing contributions $\sim 80$ times. All this indicates
that  restricting to the low energy physics not only provides the  large
distance asymptotic behavior of the vdW interaction but also the 
{\it quantitatively
dominant} contribution for {\em real graphene} for separation beyond a few
nanometers.

%\bibliography{vdw}

\begin{thebibliography}{28}
\bibitem[{wpr({\natexlab{b}})}]{wpretmet}
\bibinfo{note}{Including in-plane retardation makes the plasmon dispersion
  merge with the light-cone, but the vdW regimes of Fig.~\ref{fig:comparison}
  are still given correctly by the non-retarded plasmon expression.}
\end{thebibliography}

\end{document}